\documentclass[runningheads,a4paper]{llncs}

\usepackage{amssymb}
\usepackage{graphicx}
\usepackage{amsmath}
\usepackage{amsfonts}
\usepackage{amssymb}
\usepackage{authblk}

\usepackage{url}
\urldef{\mailsa}\path|dk3531@bristol.ac.uk|
\newcommand{\keywords}[1]{\par\addvspace\baselineskip
\noindent\keywordname\enspace\ignorespaces#1}

\begin{document}

\mainmatter  % start of an individual contribution

\title{Inferring hidden Markov models from noisy time sequences: a method to alleviate degeneracy in molecular dynamics}
%\title{Statistical Complexity of Continuous Data and Hidden Markov Modeling of Fluorescence Resonance Energy Transfer Spectra} 

\titlerunning{Inferring optimal predictors}
%\titlerunning{Statistical Complexity of Continuous Data and HMMs of FRET Spectra}

\author[1,4,*]{David Kelly}
\author[2]{Mark Dillingham}
\author[3]{Andrew Hudson}
\author[4,1]{Karoline Wiesner}

\affil[1]{Bristol Centre for Complexity Science, University of Bristol\\
Queen's Building, University Walk, Bristol, BS8 1TR, UK}
\affil[2]{School of Biochemistry, Medical Sciences Building\\
University of Bristol, University Walk, Bristol, BS8 1TD, UK}
\affil[3]{Department of Chemistry, University of Leicester\\
University Road, Leicester, LE1 7RH, UK}
\affil[4]{School of Mathematics, University of Bristol\\
University Walk, Bristol, BS8 1TW, UK}

%David Kelly, Mark Dillingham \and Karoline Wiesner}
%
\authorrunning{Inferring hidden Markov models}

\institute{}

\toctitle{Causal State Models of FRET Spectra}
\tocauthor{David Kelly, Mark Dillingham and Karoline Wiesner}
\maketitle

\begin{abstract}

We present a new method for inferring hidden Markov models from noisy time sequences without the necessity of assuming a model architecture, thus allowing for the detection of degenerate states. This is based on the statistical prediction techniques developed by Crutchfield \emph{et al.}, and generates so called causal state models, equivalent to hidden Markov models. This method is applicable to any continuous data which clusters around discrete values and exhibits multiple transitions between these values such as tethered particle motion data or Fluorescence Resonance Energy Transfer (FRET) spectra. The algorithms developed have been shown to perform well on simulated data, demonstrating the ability to recover the model used to generate the data under high noise, sparse data conditions and the ability to infer the existence of degenerate states. They have also been applied to new experimental FRET data of Holliday Junction dynamics, extracting the expected two state model and providing values for the transition rates in good agreement with previous results and with results obtained using existing maximum likelihood based methods.

\end{abstract}

\let\oldthefootnote\thefootnote
\renewcommand{\thefootnote}{\fnsymbol{footnote}}
\footnotetext[1]{To whom correspondence should be addressed. Email: \url{dk3531@bristol.ac.uk}}
\let\thefootnote\oldthefootnote

\section{Introduction}

Recent advances in experimental techniques have given new insight into many molecular systems, often on the single molecule level, see for example \cite{Hil10,Fei01,Wei00,Bus00,Sch09}. However, the data yielded from experiments at this cutting edge are frequently beset by noise which makes quantitative analysis difficult, \cite{Bro09,Kos06}. The analysis of Fluorescence Resonance Energy Transfer (FRET) spectra is a typical example of this problem.

FRET spectroscopy is a powerful method for investigating systems such as DNA molecules since it is unique in its sensitivity to molecular conformation, association, and separation in the 1-10nm range. It allows the dynamics of single molecules to be observed, avoiding the averaging inherent in ensemble measurements. In FRET spectroscopy, energy is transferred non-radiatively via a long range dipole-dipole interaction from one fluorophore to another, strategically attached to different parts of the molecule(s) under study. The efficiency of this energy transfer is strongly modulated by the separation, $R$, of the fluorophores, with a $1/R^{6}$ dependence and so is highly sensitive to changes in conformation or association. For a more detailed description of the principles and techniques of FRET spectroscopy see, for example, \cite{Ha07,Jar03} and references therein.

Since transitions between different conformational states typically take a time shorter than the resolution of the measurement, one might expect FRET spectra to exhibit jumps between discrete values (FRET efficiency levels). However, there are many sources of instrumental noise \cite{Spe08,Sis10} and also photophysical effects and temporal coarse graining. These result in the distribution of the data around some mean value, obscuring the underlying dynamics, especially in systems with many FRET levels. As the systems investigated \emph{via} FRET spectroscopy have become more complicated, a need for objective data analysis methods has been recognised. Hidden Markov Models (HMMs) are a good choice for modeling the conformational dynamics of systems. Methods of inference are well understood and the states can be interpreted as conformational states of molecules or particular associations between molecules.

However, establishing the correct model architecture (the number of states in the model and the transitions between them) is a challenge. In choosing model architecture, we must compromise between maximising the likelihood of the observations given the model and minimising the model size. It can be done using the Bayesian or Akaike Information Criteria. This is the approach taken by McKinney et al.\cite{Mck06} in prior work addressing this very problem. In this work, efficient algorithms were developed for finding model parameters which maximised the model likelihood. Then the number of states in the model was adjusted based on the average occupancy of each state, with states which were rarely visited being removed to simplify the model with only small reductions in model likelihood.

We present here an alternative method, based on statistical prediction techniques, with the same principles of maximising model likelihood and parsimony, applicable not only to FRET spectra but to any noisy time sequence displaying the following properties. Firstly the data must be clustered around discrete values. Secondly these discrete values must be sufficiently separated relative to the variance and quantity of the data (this will be explained in more detail below). Thirdly, there must be sufficient examples of switching (transitions) between these discrete values. Finally, the statistics of these transitions must be stationary, that is, the transition probabilities must be constant with time.
 
This method has the advantage that it is capable of inferring the existence of degenerate states, states associated with the same discrete value. In the context of FRET spectra, it is not necessary to associate one state with one FRET efficiency level (as is done by McKinney et al.), degenerate levels may also be discovered if revealed by the structure of the transitions between levels. In addition, the methods offer comparable performance in terms of speed and ease of use to existing model inference methods and remove the potential source of subjectivity of the selection of model architecture.

First, we will outline the theory of causal state models and the challenges to be overcome in applying such techniques to noisy time sequences. Then we shall describe the new method and the results of its application to simulated FRET spectra. Finally, we will illustrate the use of the method on the study of Holliday Junction conformational dynamics and compare this with the existing method of McKinney et al..

\section{Causal State Models}

Causal state models \cite{Sha02} are equivalent to HMMs in their structure; they both consist of a number of states connected by transitions described by a transition probability matrix and have some output (such as a real number sampled from a distribution) associated with each transition.

However, causal state models differ from HMMs in that the states represent the structure or regularities present in the data. These states are so-called \emph{causal states}; equivalence classes which group together past subsequences which share the same conditional distribution of future subsequences. In this way, if one knows what causal state a process is in, one can make as informed an estimate of the future of the process as is possible. The set of causal states is a sufficient statistic, encapsulating the same amount of information concerning the future of the process as the entire past data sequence.

To put this in more mathematical terms, let us define a bi-infinite sequence of discrete random variables representing a stationary data sequence, 
$X_{-\infty}^{\infty}=\ldots, X_{-1},X_0,X_1,\ldots$, and a particular realisation as $x_{-\infty}^{\infty}$. Then the past and future at time $t = 0$ are denoted $X_{-\infty}^{-1}= \ldots,X_{-2}, X_{-1}$ and $X_{0}^{\infty}=X_0,X_1,\ldots$ respectively and their realisations $x_{-\infty}^{-1}$ and $x_{0}^{\infty}$.

The condition of the equivalence relation is then expressed as

\begin{equation}
\epsilon(x_{-\infty}^{-1}) = \lbrace \tilde{x}^{-1}_{-\infty} : P(X_{0}^{\infty} = x_{0}^{\infty}|X_{-\infty}^{-1}=x_{-\infty}^{-1}) = P(X_{0}^{\infty} = x_{0}^{\infty}|X_{-\infty}^{-1}=\tilde{x}^{-1}_{-\infty})
\end{equation}

Note that the stationarity assumption is an important one, since the future distributions of past subsequences must be constant if we are to be able to use them for prediction.

Let $\mathcal{S}$ be the set of causal states generated from these equivalence classes. The Excess Entropy, $\mathbf{E}$, is defined as the mutual information between the past and future of the sequence and due to the sufficiency of the causal states the following is true \cite{Sha01b}.

\begin{equation}
\mathbf{E} = I[X_{-\infty}^{-1}; X_{0}^{\infty}] = I[\mathcal{S}; X_{0}^{\infty}]
\end{equation}

In the case of infinite data, a model based on causal states is provably a unique, minimal, optimal, statistical predictor of the 			future of the data sequence \cite{C01,Sha02,Sha01,Sha01b}. The proofs of the uniqueness, minimality and optimality of this statistic are out of the scope of the current work but the interested reader is referred to the original papers.

In reality, data is finite and so we must estimate the causal states based on available data. This necessitates two compromises. Firstly, the length of the past subsequences comprising the causal states must be limited such that the frequency with which the longest past subsequences are observed is sufficient to estimate the distribution of future subsequences with reasonable confidence. Secondly, the distributions of future subsequences which would be equal in the limit of infinite data will be so no longer and so a statistical test is required to determine equivalence at some chosen significance level. These practical constraints mean that there are two parameters which must be chosen, the maximum length of subsequence examined, $L$, and the test significance level, $\alpha$, although the size of the data set, $N$, and the significance level together determine the maximum length of subsequence according to the following relationship derived from entropic arguments \cite{Han93}. $|\mathcal{A}|$ is the size of the alphabet, the set of symbols present in the data.

\begin{equation}
\sqrt{\frac{|\mathcal{A}|^L}{N-L}} = \alpha
\end{equation}

Once the estimated causal states have been determined they may be linked to form an HMM by appending each of the past subsequences in the casual states with each symbol from the alphabet. The transition is determined by finding the causal state containing the resulting subsequence, with the transition probabilities determined by the relative frequencies of the new subsequences. Since the HMM must be unifilar or deterministic (the observation of a symbol when occupying a certain state must uniquely determine which state is transited to) the causal states may be split until a unifilar HMM is found. This procedure has been implemented as the Causal State Splitting Reconstruction (CSSR) algorithm by Shalizi and Klinkner and is described in \cite{Sha04}.

Causal State models have been applied to many systems including spin systems \cite{C31}, crystal growth \cite{Var02}, molecular dynamics \cite{Li08}, atmospheric turbulence \cite{Pal02}, population dynamics \cite{Cru06,Gor08}, self-organisation \cite{Sha04} and neural spike sequences \cite{Tin02}.

\subsection{Application to FRET Spectra}

Data in the real world is rarely discrete. The discrete data upon which these methods are based is assumed to have been observed via some measurement channel with a finite resolution designated $\epsilon$. Obviously, the HMM obtained is strongly dependent on this resolution and the HMM is also referred to as an $\epsilon$-machine to emphasise this dependence.

If we are to apply these methods to FRET spectroscopy, we wish our resulting HMM to be independent of the discretisation scheme used to obtain it, since for the model to be useful it should be determined by the underlying system, not by the particulars of the method used to obtain it.

FRET spectra would ideally be discrete since the system undergoes transitions between conformational states corresponding to certain FRET efficiencies on a timescale shorter than that of observations, resulting in discrete jumps between FRET levels. It is a natural choice, therefore, to base any discretisation scheme on these FRET levels.

However, there are many experimental sources of noise which result in data being $\beta$-distributed (or to a reasonable approximation normally distributed) around the idealised FRET levels with distributions typically overlapping \cite{Dah99}. This noise in (simulated) spectra makes it impossible to determine with certainty to which FRET level each data point should belong. Misassignment of FRET levels distorts distributions and introduces fallacious structure which, in the case of simulated data, leads to inferred HMMs varying from the models used to generate the data.

The methods presented in the next section address this problem, allowing the underlying model to be identified.

\section{Methods}

In contrast to conventional methods (which typically ignore uncertainty in assignments), explicitly recognising uncertainty in the discretisation allows the problem of noise to be circumvented. By assigning a special null symbol to any data point which could not be reliably assigned to a FRET level and then disregarding these symbols when determining causal states, the underlying model architecture (that used to generate the data in the case of simulated spectra) could be inferred.

\begin{figure}
\centering
\includegraphics[width=0.9\textwidth]{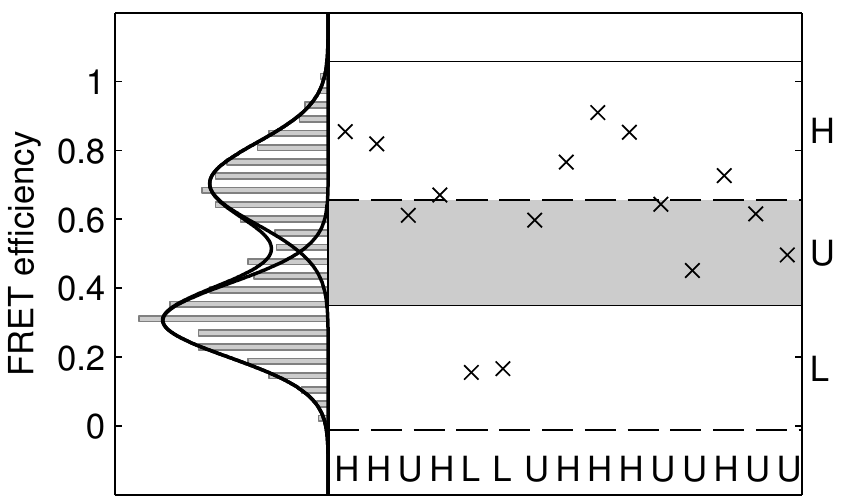}
\caption{On the vertical axis the histogram of the spectrum is shown, along with the fitted Gaussian mixture model. The resulting partitions are shown with solid horizontal lines where the upper component's probability reaches 0.001 and dashed lines for the lower component. A short section of the spectrum is also shown with the corresponding symbol sequence. Here H and L correspond to the high and low FRET levels and U indicates uncertainty.}
\label{fig:1}
\end{figure}

The procedure (illustrated in Fig.~\ref{fig:1}) is as follows;

\begin{enumerate}
\item Construct a histogram of FRET efficiencies.
\item Fit Gaussian mixture models with varying numbers of components.
\item Select a mixture model using the Akaike Information Criterion.
\item Partition the space. For a model with $n$ components there will be $2n$ partition boundaries, located where the probability density of each model component reaches some small, user defined limit (i.e. 0.001). There will be $2n-1$ bounded regions defined by these boundaries.

The partition boundaries associated with each model component may or may not overlap with partitions associated with other model components depending on the separation of the means relative to the variances. In either case the odd numbered regions correspond to certain assignment of data points to one model component. The even numbered regions in between correspond to regions of uncertainty. Here there is a non-negligible probability of a data point being generated by more than one model component, either because model components overlap or because the probability of a data point being generated by any component is very low.

Note that this partitioning assumes that the partitions associated with any one model component do not both fall in between the partitions associated with another, an unlikely circumstance which could only occur with FRET levels extremely close together or with very different variances. If this does occur, appropriate partitions cannot be found.

\item Calculate the ratio of the area of the partition to the area of the model component with which it is associated for each partition-model component pair. Find the minimum of these ratios and adjust the partitions of the other areas in order to equalise them.
The reason for this is that this partitioning effectively discards a proportion of the occurrences of each possible subsequence in the discretised data. If we discard more of one subsequence than another we skew their relative frequencies and, as a result, alter the transition probabilities of the HMM. By maintaining the original ratios between model components in the partitioning we avoid this source of bias.

\item Assign each data point a symbol based on the partition in which it lies. Points which were generated by one component of the mixture model with high probability ($> 0.999$) are assigned the symbol corresponding to this component. Points located where there is any overlap of components are assigned the null symbol.
\item Determine the causal states of the model using an adapted version of the CSSR algorithm. The adaptation is to only append symbols which are certain to existing subsequences (starting with the empty subsequence) so subsequences containing the null symbol are never considered.
\end{enumerate}

Since Gaussian distributions have a non-zero probability of generating a data point any distance from their mean, there is still a small probability of misassignment of data points. If this occurs there may be extra transitions present in the inferred HMM, however the probability of these transitions is generally very small relative to other transitions present and as such may be easily identified. There is necessarily a compromise between obtaining a sufficient proportion of non-null symbols to be able to determine the causal states and avoiding misassignment. The location of the partitions with regards to this compromise will be dictated by the data; it is easier to avoid misassignments where the FRET levels are widely spaced. These methods have been implemented in \textsc{Matlab}.

\section{Simulated Data}

We demonstrate the algorithm with simulated FRET data. A typical FRET system was simulated using the HMM shown in Fig.~\ref{fig:2}. Rather than outputting a particular symbol on each transition, a Gaussian function, $f_0$ or $f_1$, was sampled. The means of the two functions were 0.3 and 0.7 and the standard deviation was 0.1 for both. The length of the data series was 1500. The fit of the Gaussian mixture model to the histogram is shown in Fig.~\ref{fig:1} along with the partitions and a small portion of the spectrum to demonstrate the symbolisation.

\subsection{Results}

\begin{figure}
\centering
\includegraphics[width=0.5\textwidth, trim = 20mm 146mm 194mm 9mm, clip]{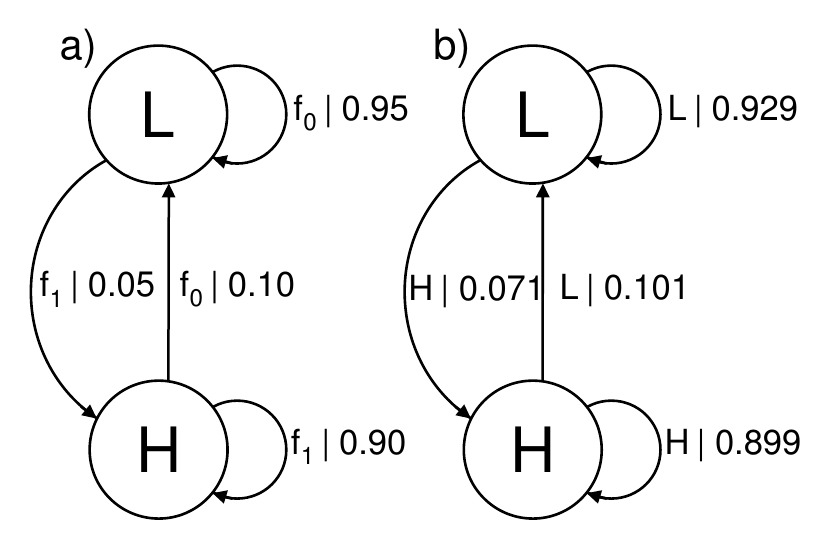}
\caption{a) The HMM used to generate the data and b) the HMM inferred from the data. For the generating model the transitions are labelled with the function sampled to generate a data point and its probability. For the inferred model the transitions are labelled with the symbol output on the transition and its probability.}
\label{fig:2}
\end{figure}

A typical example of a HMM inferred from the symbolised data is shown in Fig.~\ref{fig:2}. As can be seen, the generating and inferred model are very similar, with the correct architecture being inferred. To quantify this let us define the model distance, following Rabiner \cite{Rab89}, as the difference in the log probabilities of the observed data, $O^{(2)}$, being generated by the generating model and the inferred model, designated $\lambda_2$ and $\lambda_1$ respectively, normalised for the length of the data, $N$.

\begin{equation}
D(\lambda_1, \lambda_2) = \frac{1}{N} ( \log P(O^{(2)}|\lambda_1) - \log P(O^{(2)}|\lambda_2))
\end{equation}

This measure is equal to zero for models with the same statistical properties and in this case the model distance is close to zero, 0.016, averaged over 5 repetitions, with a standard deviation of 0.009. The small error is due to the difficulty in estimating the exact distributions with data sets of this size. The methods are, therefore, capable of inferring accurate models under conditions typical to real data.

\subsection{Degenerate systems}

\begin{figure}
\centering
\includegraphics[width=0.5\textwidth, trim = 20mm 98mm 190mm 9mm, clip]{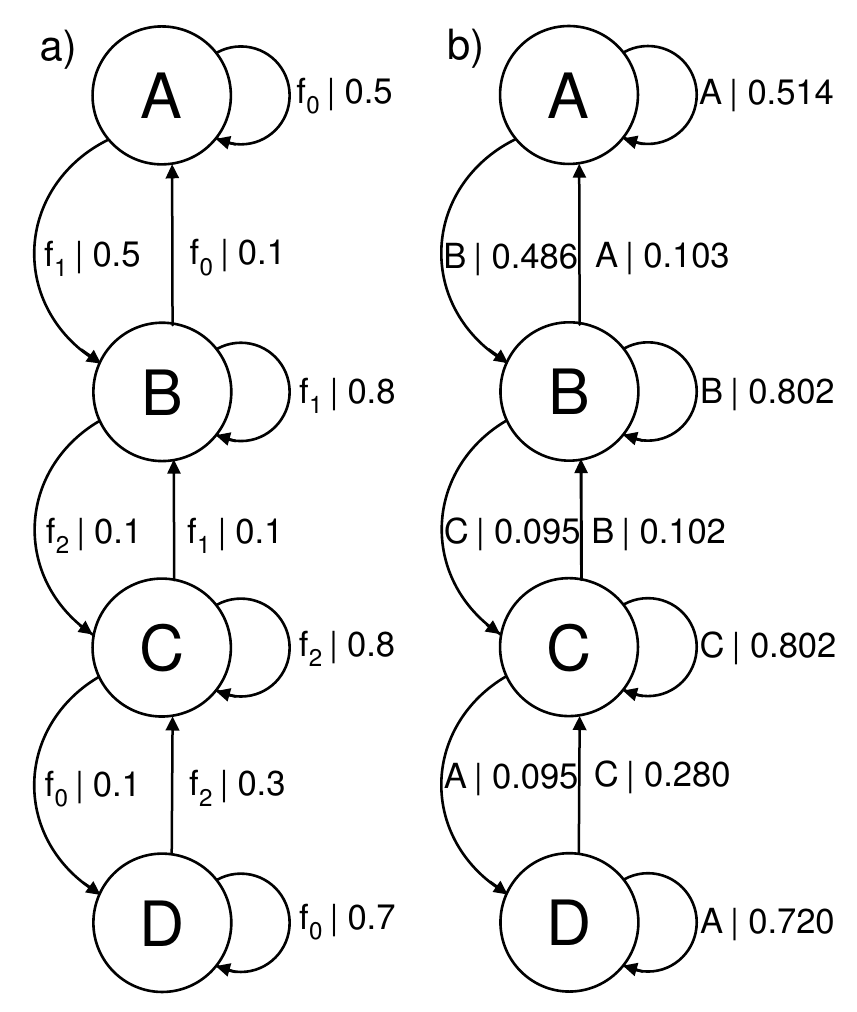}
\caption{a) Model used to generate the data. This 4 state model has two states associated with the FRET level centred at 0.1 (denoted $f_0$) but with different probabilities of remaining in each state. b) The model inferred from the data.}
\label{fig:3}
\end{figure}

\begin{figure}
\centering
\includegraphics[width=0.5\textwidth, trim = 20mm 98mm 190mm 9mm, clip]{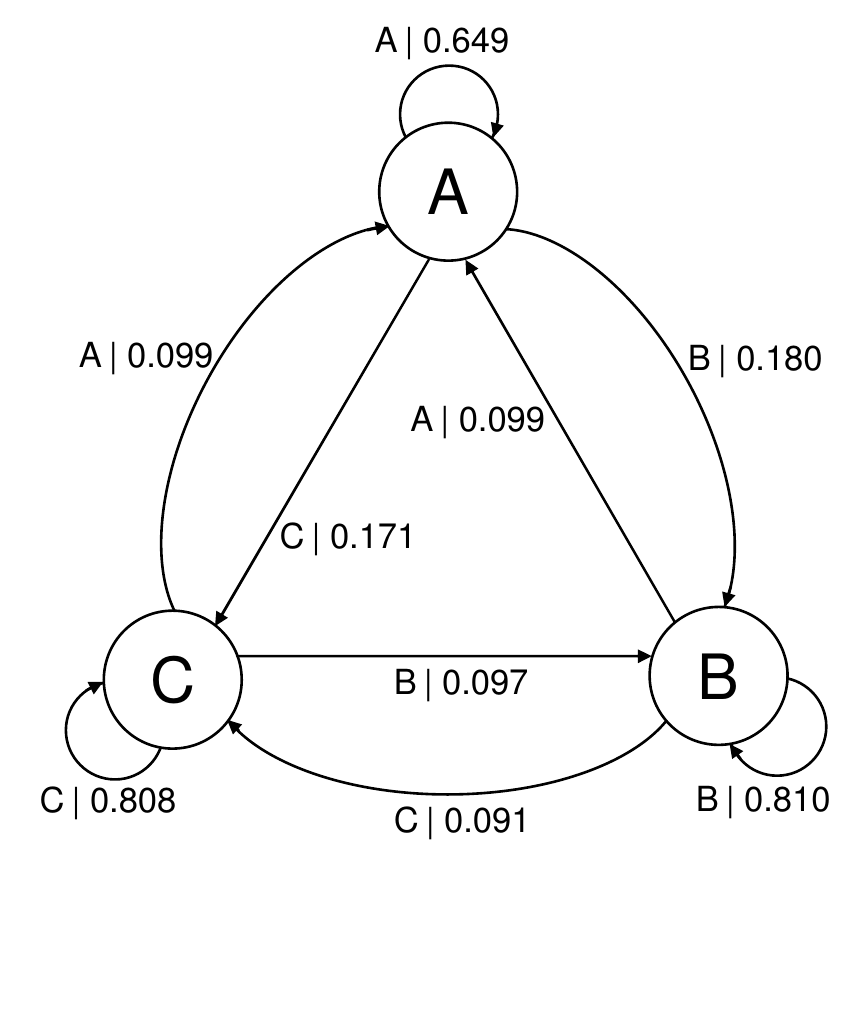}
\caption{HaMMy cannot distinguish between the two degenerate states (A and D in Fig.~3a) resulting in a model with a state (labelled A) averaging the degenerate states' transition probabilities.}
\label{fig:4}
\end{figure}

To demonstrate the ability of the methods to identify structure in data where different hidden states are associated with the same observable - degenerate systems - we also simulated data using the model shown in Fig.~\ref{fig:3}. Since this system is more complicated the data requirements to infer the correct architecture are comparatively higher; the result (also shown in Fig.~\ref{fig:3}) was obtained for 5000 data points. The Gaussian functions sampled on the transitions had means of 0.1, 0.5 and 0.9 and standard deviations of 0.09. In comparison, existing methods for inferring hidden Markov models from FRET data such as HaMMy, described in more detail below, may only hope to extract a 3 state model due to the constraint of associating each FRET level with one state. The `HaMMy' programme was also run on this spectrum obtaining the 3 state model shown in Fig.~\ref{fig:4}. Note however, that one could identify states that had multiple transition rates associated with them by plotting histograms of the dwell times in each state as in the work by Laurens et al. \cite{Lau09}.

\section{Experimental Data}

\begin{figure}
\centering
\includegraphics[width=0.5\textwidth, trim = 20mm 127mm 190mm 9mm, clip]{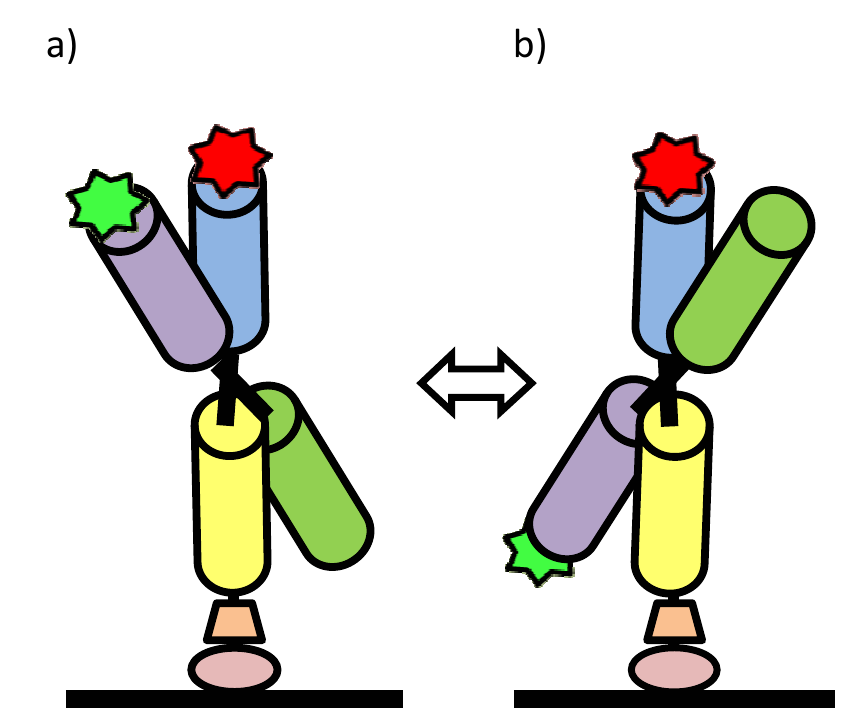}
\caption{Schematic of the two stable `Stacked X' conformers of a Holliday junction. The four cylinders represent the arms of the junction and the red and green stars represent the attached fluorophores. The junctions can interchange between conformers and in doing so go from the high FRET (a) to the low FRET (b) state (or vice versa). The pink and orange shapes represent the streptavidin - biotinylated BSA bridge used to specifically bind the junctions to the slide.}
\label{fig:5}
\end{figure}

Holliday Junctions are cross shaped, four way junctions of DNA and important intermediates in DNA recombination. As such they have been studied extensively \cite{Yu04,Mck04,Mck05,Mck06}. In the presence of divalent metal ions such as Mg$^{2+}$ they have two stable conformations known as `stacked X' conformers illustrated in Fig.~\ref{fig:5}. Junctions will switch stochastically between the two conformations at a rate determined by the concentration of magnesium ions. If fluorescent probes are attached to the arms of the junction then these conformational changes may be observed by a change in FRET efficiency. Prior work has identified DNA sequences which form Holliday junctions with an approximately equal occupation of each conformer and characterised the dependency of the transition rate on the concentration of magnesium ions \cite{Mck03}. In order to test the methods on experimental data, these experiments were repeated and causal state models were successfully constructed from the resulting data.

\subsection{Experimental Methods}

Biotin-labelled Holliday junctions (identical to `Junction 7') were assembled and purified essentially according to published methods (McKinney et al. \cite{Mck03}). Equivalent junctions without donor and/or acceptor fluorophores were prepared in the same manner for use as controls. The junctions with only one fluorophore are used for collecting data with which to correct the FRET efficiency for overlap of the emission spectra of the two fluorophores. The junctions with no fluorophores are used to confirm a low level of background fluorescent contaminants. The junctions were bound to a cover glass (Menzel Glaser Nr 1.5) with a BSA-biotin streptavidin bridge using a modification of the method of McKinney et al.. Briefly, the cover glass was cleaned with an argon plasma, then treated with biotinylated BSA (1 mg/ml, Sigma) for 5 minutes before washing extensively with T50 buffer (10 mM Tris-HCL [pH 7.5], 50mM NaCl). Streptavidin (0.2 mg/ml, Invitrogen) was applied for 2 minutes before washing as before. A four channel imaging cell was constructed by sandwiching appropriately cut double-sided tape between the modified cover glass and a plasma-cleaned microscope slide. Holliday junctions (50 pM molecules) were added to the channel and incubated for 5 minutes before washing with T50 buffer supplemented with MgCl$_2$ (as stated), an oxygen scavenger system (1 mg/ml glucose oxidase, 0.04 mg/ml catalase, and 0.8 mg/ml dextrose, Sigma) and anti-photobleaching reagents (1 mM methylviologen, 1 mM Ascorbic Acid, Sigma)\cite{Vog08}.

\begin{figure}
\centering
\includegraphics[width=0.5\textwidth, trim = 20mm 92mm 102mm 9mm, clip]{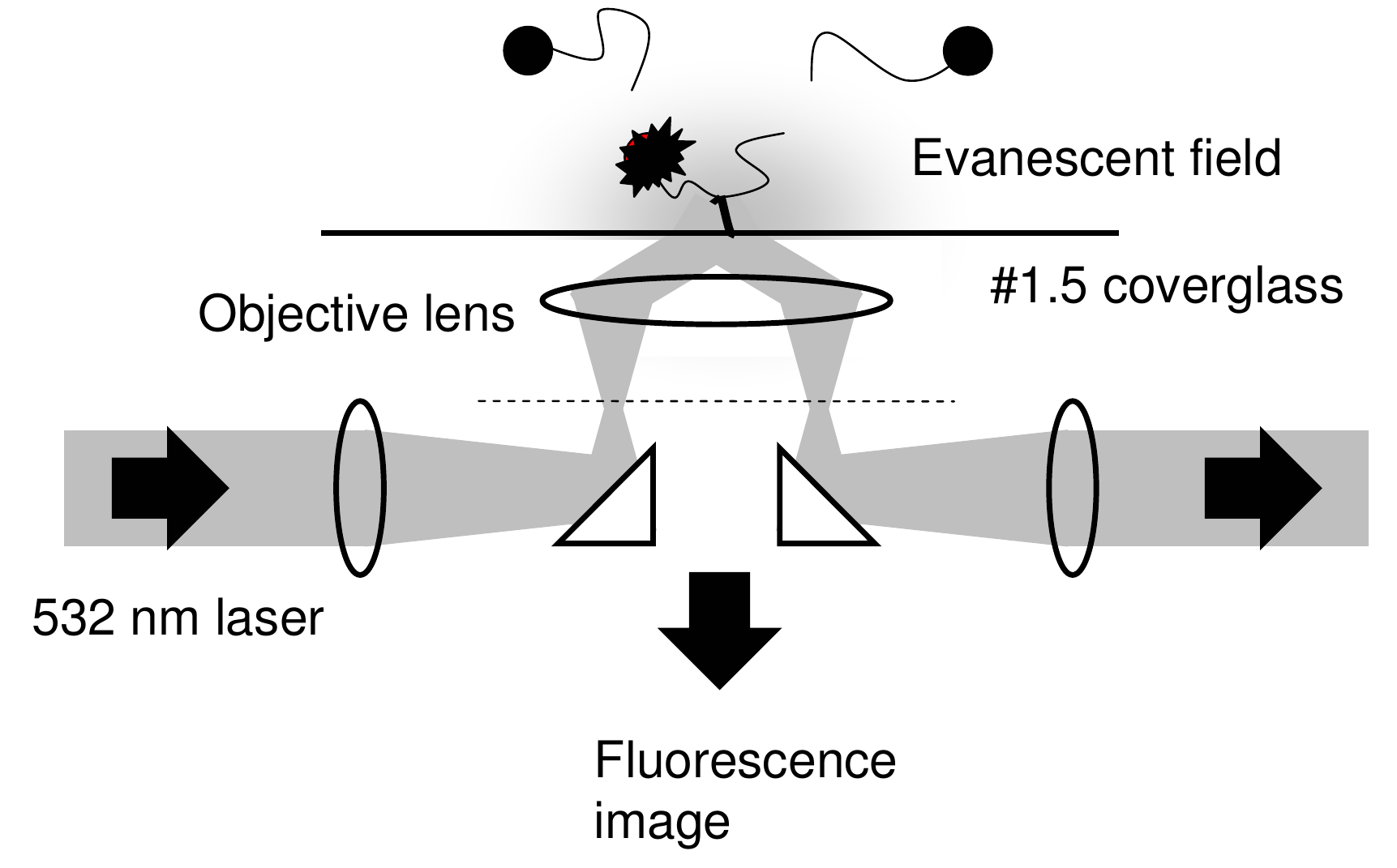}
\caption{Schematic of the optical design for TIRF illumination }
\label{fig:6}
\end{figure}

FRET spectra were obtained using a custom built objective-based total-internal-reflection fluorescence (TIRF) microscope which is very similar in design to one described in detail elsewhere \cite{Mas07}. A schematic is shown in Fig.~\ref{fig:6}. Excitation was achieved using a 100 mW 532 nm laser (Laser Quantum, Ventus) attenuated by neutral-density filters. Emission light passed through a 532 nm notch filter (Semrock, StopLine) to remove scattered laser light and then a commercial dual-view system (Optosplit II, Cairn) to produce two images corresponding to the fluorescence from Cy3 (bandpass filter centred at 580 nm, width 60 nm) and Cy5 (bandpass filter centred at 655 nm, width 65nm). Images were recorded using an electron-multiplied charge-coupled device (EM-CCD, iXon Du 897, Andor Technologies) with the Solis software package (Andor Technologies). For each dataset, the brightest objects were identified in each channel, matched between channels and the intensity timeseries extracted. Where these timeseries showed anticorrelation over a long period, FRET efficiencies were calculated according to methods in \cite{Ha07} which includes a correction for leakage of the Cy3 emission into the Cy5 channel. Each FRET spectrum was then discretised using the methods described above and passed to the CSSR algorithm to construct causal state models. The models were then used with the transition probabilities to calculate the average transition rate for the junctions for each concentration. The spectra were also analysed using the `HaMMy' programme as described \cite{HaMMy}.

\subsection{Results}

Thirty nine spectra of varying lengths were obtained for a range of different magnesium ion concentrations. These were analysed with the HaMMy programme and with the new methods described above.

\subsubsection{HaMMy Results}

Briefly, the HaMMY programme works in the following way, for more detail the reader is referred to the original paper \cite{Mck06} and references therein. First the user specifies the number of states (FRET levels) they wish to fit to the data. This determines the number of parameters in the model. These parameters are then varied in order to maximise the likelihood of observing the data using Brent's algorithm, a multi-dimensional optimisation algorithm. At each step in Brent's algorithm, i.e. for each set of parameter values, the likelihood of the data is calculated using the Viterbi algorithm (an efficient method, guaranteed to find the most probable state sequence). Providing the procedure does not converge to a local maximum rather than the global maximum it should infer the model with maximum likelihood of generating the data. Then one can examine the fitted spectrum and identify and eliminate extraneous states if they are never, or very infrequently, visited. Since we may identify and remove extraneous states but not add more, it is prudent when initially specifying the number of states to overestimate (by two as a rule of thumb).

Following this, the programme was run first of all with four states. Frequently this resulted in three FRET levels being visited in the idealised spectrum, two FRET levels very close together where one would assume there was only one, a case of the algorithm converging to a local maximum since the initial conditions were such that two FRET levels were equidistant from the actual FRET level and so both converged upon it. To circumvent this problem, initial guesses were supplied to the algorithm close to the actual FRET levels. The remaining spectra were fitted in this way. The HaMMy programme was able to infer a two state model for all of the spectra; extra states were hardly ever visited and for the most part had unphysical FRET values greater than one.

\subsubsection{Causal State Modelling Results}

The Causal State Modelling algorithms were also run on the data. It was found that although the requirements of the data for these methods were more stringent they could be successfully applied in the majority of cases.

The parameters of the inference algorithm were determined as follows. The significance level for the statistical test was set at 0.05. Then entropic considerations as to the likelihood of statistical fluctuations significant at this level guide an appropriate choice of maximum subsequence length. Since in these spectra data are relatively scarce, especially if the spacing of the FRET levels means a low percentage of the data are used, the maximum subsequence length was typically low, specifically 2. For longer spectra this was increased where possible. In seven cases, the number of data were insufficient to justify even this length of subsequence. However, 2 was still used in these cases, as to reduce it further would constrain the algorithm to a maximum of two states and so be an unfair comparison.

Two state models were inferred for thirty of the thirty nine spectra. Of those that failed seven were due to the FRET levels being too close together. In these cases, there were insufficient `certain' data after the discretisation to be able to infer a model. Of these seven, in two borderline cases a model was inferred but the transition architecture was incorrect. In the remaining two cases the failure was due to the FRET levels changing monotonically with time so as to cross the partitions meaning no transitions between `certain' symbols could be observed and hence no model inferred.

It was also found that, due to the high level of noise and the slight changes of FRET level with time leading to a higher weight between the two peaks, the routine often inferred a mixture model with more than two components despite the histogram of the FRET efficiencies clearly having two peaks. This may also have been due in part to the integration time of the camera averaging over transitions between states. In these cases, where two components were a more appropriate representation, the routine was constrained to fit the mixture model as such. Note that this constraint has no bearing on the number of states in the HMM which is still unconstrained.

\begin{figure}
\centering
\includegraphics[width=0.5\textwidth]{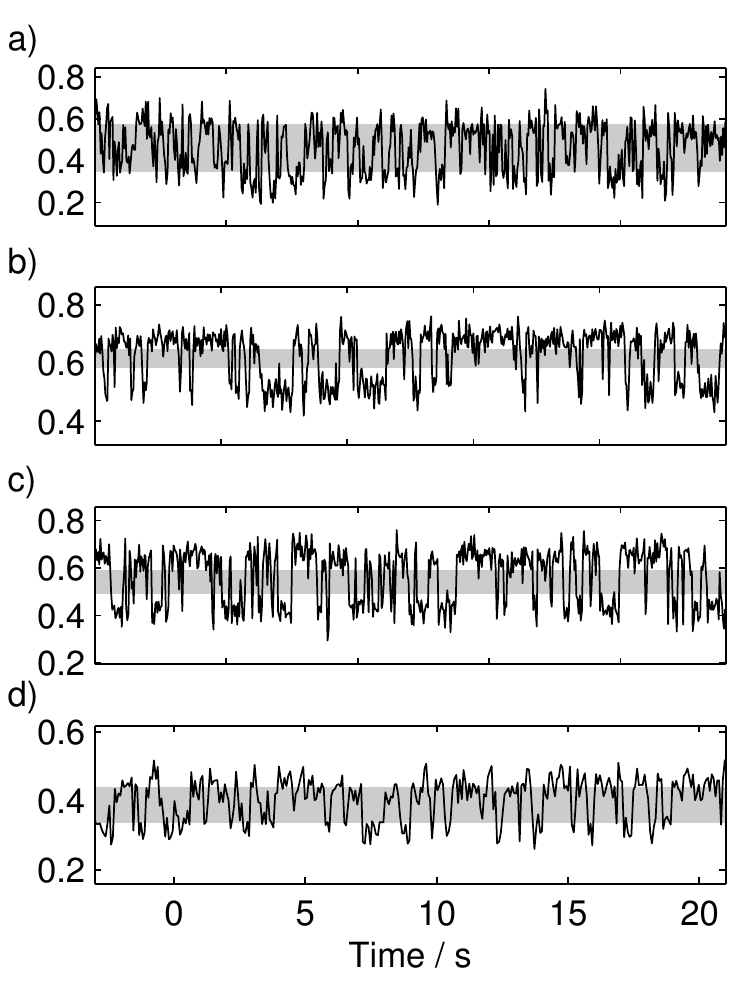}
\caption{Example sections of spectra for Mg$^{2+}$ concentrations a) 30 mM, b) 40 mM, c) 50 mM and d) 60 mM. The shaded region corresponds to the uncertain partition.}
\label{fig:7}
\end{figure}

\begin{figure}
\centering
\includegraphics[width=0.5\textwidth, trim = 20mm 98mm 190mm 9mm, clip]{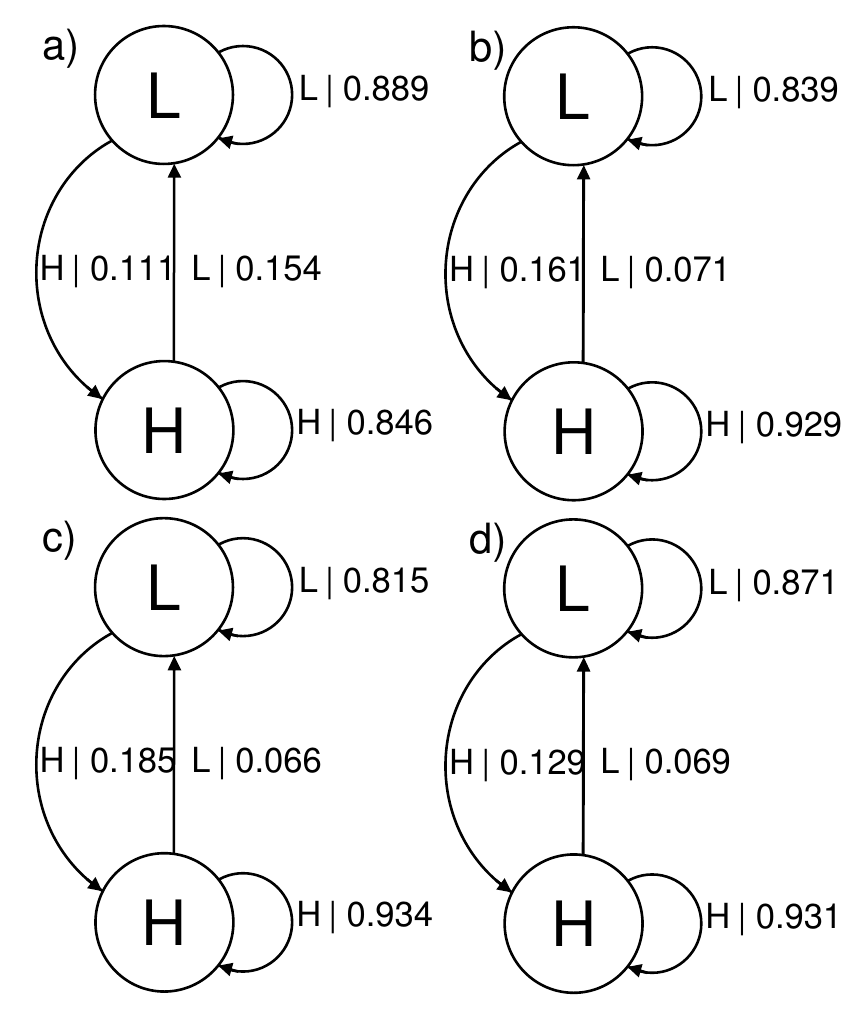}
\caption{Causal state machines corresponding to the 4 spectra shown above with Mg$^{2+}$ concentrations a) 30 mM, b) 40 mM, c) 50 mM and d) 60 mM . Note that the atual transition rates are given by dividing transition probabilities by the sampling rate of the data, these were 41 ms per point for 30 - 50 mM and 71 ms per point for 60mM.}
\label{fig:8}
\end{figure}

Despite the problems outlined above, the methods performed well for the less noisy spectra of reasonable length. In Fig.~\ref{fig:7} some example spectra are shown along with the resultant causal state models in Fig.~\ref{fig:8}.

\section{Method Comparison}

The two methods are both capable of inferring models in agreement with our intuition and understanding of the physical system generating the data, but make different assumptions and have different requirements of the data.

HaMMy requires as an input the number of FRET levels the user believes are present in the spectrum (over estimated to ensure the procedure is not constrained to fit a sub-optimal model) and assumes a model architecture with a state corresponding to each FRET level. Additional inputs specifying initial parameter values close to true values may improve the performance of the algorithm.

The causal state methods require (in the case of noisy data) the number of FRET levels the user believes are present, and a significance level at which to test whether  or not distributions are equivalent. This significance level along with the quantity of data determines the remaining parameter, the maximum length of subsequence examined. The causal state methods make no assumptions regarding the model architecture but increase the number of states in the model if the current model cannot adequately account for structure in the data. They also allow for degeneracy, more than one state associated with the same FRET level. Both methods assume stationarity. As seen from the results above, the causal state methods have more stringent requirements regarding the quantity and quality of data.

\begin{figure}
\centering
\includegraphics[width=0.5\textwidth]{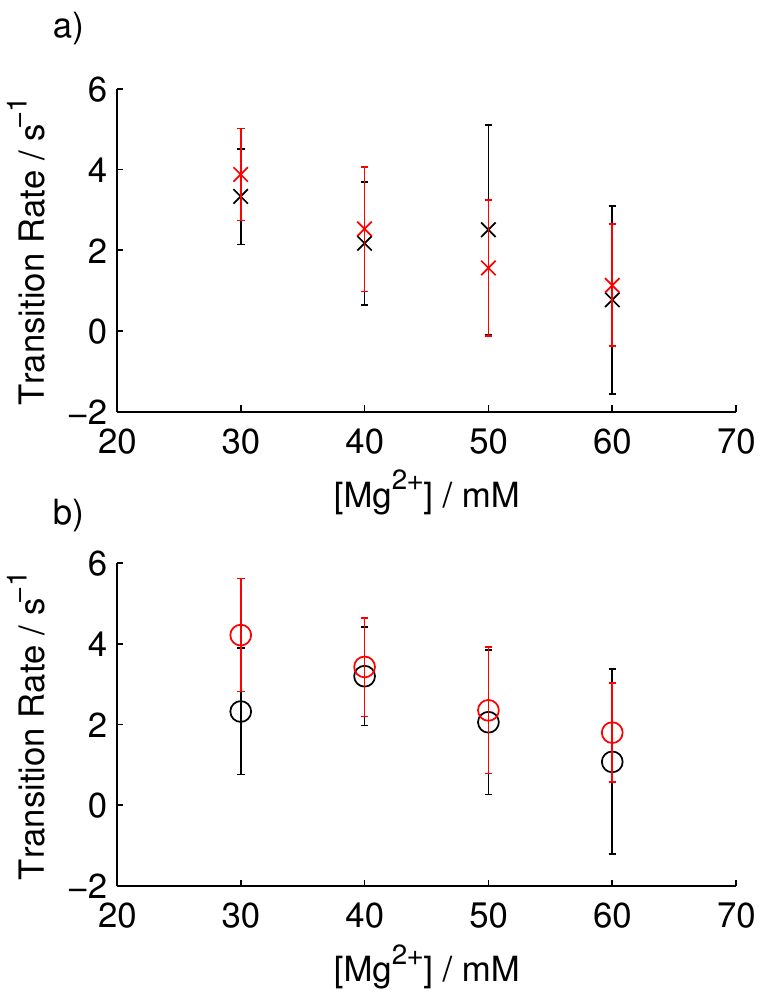}
\caption{Average transition rates from a) the high FRET state to the low FRET state and b) the low FRET state to the high FRET state as a function of magnesium ion concentration for HaMMy (red) and the causal state method (black). The error bars indicate the standard deviation.}
\label{fig:9}
\end{figure}

The transition rates as a function of Mg$^{2+}$ concentration are shown in Fig.~\ref{fig:9} for both analysis methods. Note that these values are average results for multiple spectra, obtained by taking logs, calculating the mean and standard deviation for these transformed values, then exponentiating as in \cite{Mck06}. These values are in good agreement with previous work \cite{Mck03}, exhibiting the same trend and being of the same order of magnitude; exact values for transition rates may vary with temperature. The values from the two different methods are consistent with each other in that the differences between them are within the error tolerances, however, we observe that the results from CSSR are consistently lower than those from HaMMY. We believe this is due to the causal state modelling underestimating the transition probabilities for the following reason. Since the data are time binned, all transitions must occur within an integration period resulting in a value of FRET efficiency for that bin which has been averaged to some extent. Due to the partitioning and discretisation scheme, these time averaged bins are more likely to be discounted by the causal state inference algorithm since they are more likely to fall in the ambiguous region between the two peaks in FRET efficiency. This introduces a bias into the statistics since time bins containing no transitions are less likely to be discounted in this way. For high data sampling rates relative to the time scale upon which the transitions occur this bias will be negligible, however, if the sampling rate is too low then the bias will become significant, as is the case for the rate inferred for the 30 mM magnesium ion concentration data. Since the simulated data was not subjected to further sampling or coarse graining this biasing was not observed and the correct transition probabilities were inferred.

\section{Conclusions and Future Work}

This paper presents a new method for inferring Hidden Markov models from noisy time series, demonstrating the ability to infer the correct model architecture with minimal initial assumptions. We emphasise that the method is not only applicable to FRET spectra, but to any data source with a natural tendency to cluster. It will generate unique, optimal and minimal predictors with only 2 input parameters. Application to the conformational dynamics of Holliday Junctions has demonstrated the ability of the methods to extract models from experimental data which agree with previous work in both model architecture and transition rates. The method provides a complementary alternative to existing methods of fitting HMMs to FRET spectra. Comparison between the new method and an existing maximum likelihood method shows that the requirements for the new method are more stringent; requiring a sufficient spacing of FRET levels, a sufficient quantity of data and a high sampling rate relative to the timescale of the dynamics of interest. However, since this new technique should be able to directly discern multiple states with the same FRET distribution it holds a considerable advantage over its predecessor. Future work will utilise these abilities to examine more complicated, possibly degenerate systems, such as RecA nucleoprotein filament formation on DNA, to provide greater insight than traditional analysis methods.

\medskip %adds vertical space apparently

\bibliographystyle{unsrt}
\bibliography{InferringHMMs}

\end{document}